\documentclass[aps,prl,twocolumn,superscriptaddress, preprintnumbers]{revtex4-1}

\usepackage{units}
\usepackage{upgreek}
\usepackage{amsmath,amssymb}
\usepackage{subscript}
\usepackage{graphicx}

\usepackage{xcolor}
\definecolor{darkred}{rgb}{0.5,0,0}
\definecolor{darkblue}{rgb}{0,0,0.5}
\definecolor{firebrick}{rgb}{0.75,0.125,0.125}
\definecolor{darkgreen}{rgb}{0,0.5,0}
\usepackage[colorlinks=true,linkcolor=firebrick,citecolor=darkgreen,urlcolor=darkblue]{hyperref} 

\usepackage{xspace}

\begin{document}

\preprint{Published in Phys. Rev. Lett. as DOI: \href{http://dx.doi.org/10.1103/PhysRevLett.116.241101}{10.1103/PhysRevLett.116.241101}}

\title{Measurement of the Radiation Energy in the Radio Signal of 
Extensive Air Showers as a Universal Estimator of Cosmic-Ray Energy}

\author{A.~Aab}
\affiliation{Universit\"{a}t Siegen, Fachbereich 7 Physik - 
Experimentelle Teilchenphysik, Siegen, 
Germany}
\author{P.~Abreu}
\affiliation{Laborat\'{o}rio de Instrumenta\c{c}\~{a}o e F\'{\i}sica Experimental 
de Part\'{\i}culas - LIP and  Instituto Superior T\'{e}cnico - IST, 
Universidade de Lisboa - UL, Lisboa, 
Portugal}
\author{M.~Aglietta}
\affiliation{Osservatorio Astrofisico di Torino  (INAF), 
Universit\`{a} di Torino and Sezione INFN, Torino, 
Italy}
\author{E.J.~Ahn}
\affiliation{Fermilab, Batavia, IL, 
USA}
\author{I.~Al Samarai}
\affiliation{Institut de Physique Nucl\'{e}aire d'Orsay (IPNO), 
Universit\'{e} Paris 11, CNRS-IN2P3, Orsay, 
France}
\author{I.F.M.~Albuquerque}
\affiliation{Universidade de S\~{a}o Paulo, Instituto de F\'{\i}sica, S\~{a}o 
Paulo, SP, 
Brazil}
\author{I.~Allekotte}
\affiliation{Centro At\'{o}mico Bariloche and Instituto Balseiro 
(CNEA-UNCuyo-CONICET), San Carlos de Bariloche, 
Argentina}
\author{P.~Allison}
\affiliation{Ohio State University, Columbus, OH, 
USA}
\author{A.~Almela}
\affiliation{Universidad Tecnol\'{o}gica Nacional - Facultad Regional 
Buenos Aires, Buenos Aires, 
Argentina}
\affiliation{Instituto de Tecnolog\'{\i}as en Detecci\'{o}n y 
Astropart\'{\i}culas (CNEA, CONICET, UNSAM), Buenos Aires, 
Argentina}
\author{J.~Alvarez Castillo}
\affiliation{Universidad Nacional Aut\'{o}noma de M\'{e}xico, M\'{e}xico, 
D.F., 
M\'{e}xico}
% ...
\collaboration{The Pierre Auger Collaboration}
\email{{\tt auger\_spokespersons@fnal.gov}}
\noaffiliation

\begin{abstract}
We measure the energy emitted by extensive air showers in
the form of radio emission in the frequency 
range from 30 to \unit[80]{MHz}. Exploiting the accurate 
energy scale of the Pierre Auger Observatory, we obtain a
\emph{radiation energy} of \unit[$15.8 \pm 0.7\,\mathrm{(stat)} \pm 6.7\,
\mathrm{(sys)}$]{MeV} for cosmic rays with an energy of 
\unit[1]{EeV} arriving perpendicularly to a geomagnetic field of 
\unit[0.24]{G}, scaling quadratically with the cosmic-ray energy. A 
comparison with predictions from state-of-the-art first-principle 
calculations shows agreement with our measurement.
The radiation energy provides direct access to the 
calorimetric energy in the electromagnetic cascade of extensive air 
showers. Comparison with our result thus allows the direct calibration
of any cosmic-ray radio detector against the well-established energy
scale of the Pierre Auger Observatory.
\end{abstract}


\maketitle



\label{sec:introduction}In this work,
we address one of the most important 
challenges in cosmic-ray physics: the accurate determination of the
absolute energy scale of cosmic rays. Measurements with surface 
particle detector arrays rely on assumptions about cosmic-ray 
composition and on extrapolations of our knowledge about 
hadronic interactions to energies beyond the reach of the Large Hadron 
Collider. Consequently, their determination of the absolute cosmic-ray energy suffers from significant uncertainties \cite{Engel:2011zzb}.
Fluorescence detectors measure the calorimetric energy in 
the electromagnetic cascade of air showers, which allows 
an accurate determination of the energy of the primary particle \cite{EnergyScaleICRC2013}.
However, fluorescence light detection is only possible at sites with 
good atmospheric conditions, and precise quantification of scattering and absorption of fluorescence light under
changing atmospheric conditions requires extensive atmospheric
monitoring efforts \cite{Abreu:2012oza,Abraham:2010pf,Abraham:2009bc,Abreu:2013qtw}.

An attractive option to determine the energy scale of cosmic-ray 
particles is given by the detection of radio signals. Radio detection 
of extensive air showers can be performed at any site not overwhelmed 
by anthropogenic radio signals, requiring only detector arrays of moderate size and complexity.
It has been known since the 1960s that air showers emit measurable 
radio pulses \cite{Allan1971}. The physics of the radio emission from extensive air showers is by
now well understood (see \cite{HuegeRadioReview2016} for an overview).
The radiation dominantly arises from geomagnetically induced, time-varying transverse 
currents \cite{KahnLerche1966,Werner2008} and is strongly
forward beamed in a cone of a few degree opening angle due to the 
relativistic speed of the emitting particles. The atmosphere is 
transparent for radio waves at the relevant frequencies, i.e., scattering and absorption are 
negligible. As the emission is generally coherent at frequencies below \unit[100]{MHz}, the amplitude of the electric 
field scales linearly with the number of electrons and positrons in 
the air-shower cascade, which in turn scales linearly with the primary 
cosmic-ray energy.

Several analyses exploiting this calorimetric property of the radio emission for the determination of the 
energy of cosmic-ray particles have previously been published 
\cite{LOPES2005,GlaserARENA2012,LOPES_energyxmax_2014,Bezyazeekov201589}.
All of these approaches used the radio-signal strength at
a characteristic lateral distance from the shower axis as an estimator for the
cosmic-ray energy. While this method has long been known to provide good precision 
\cite{Huege2008}, it has the marked disadvantage that the corresponding 
energy estimator cannot be directly compared across different experiments. Asymmetries arising from 
the charge-excess contribution 
\cite{Askaryan1962,Marin2011,AERAPolarization} can be corrected for, and the air-shower 
zenith angle can be normalized out. The systematic influence of the 
observation altitude on the lateral signal distribution, however, poses a 
fundamental problem for such comparisons. In a simulation study, we have quantified the 
difference between radio amplitudes at the characteristic lateral 
distance measured for the same showers at sea level (altitude of LOFAR \cite{LOFARInstrument}) and at 
\unit[1560]{m} above sea level (altitude of the radio detector array 
of the Pierre Auger Observatory \cite{Auger2014}). We observe 
differences between -11\% and +23\% with an average deviation of 11\%.
These deviations in the measured amplitude arise from the
fact that the lateral radio signal distribution flattens 
systematically with increasing distance of the radio antennas to the 
air-shower maximum. Furthermore, the optimal lateral distance at which to make the measurement also varies with observation altitude \cite{PRLSupplement}.
While absolute values for the amplitudes measured at a characteristic lateral
distance as a function of cosmic-ray energy have been published by several experiments
\cite{LOPES_energyxmax_2014,PHDWeidenhaupt,Bezyazeekov201589}, no direct
comparison between the energy scales of these cosmic-ray radio detectors
has therefore been performed to date. (Most experiments obtain their 
energy scale based on surface detector arrays and thus incur 
uncertainties from hadronic interaction models.)

Here, we make an important conceptual step forward in using 
radio signals from extensive air showers for the absolute calibration of the energy scale of cosmic-ray 
detectors. We use the total energy radiated by extensive air showers in the form of 
radio emission in the frequency range from 30 to \unit[80]{MHz}, 
hereafter called \emph{radiation energy}, as an estimator of the 
cosmic-ray energy. Due to conservation of energy, and the absence of 
absorption in the atmosphere, the radiation energy measured at 
different observation altitudes is virtually identical. In the 
above-mentioned simulation study, the radiation energy was shown to 
vary less than 0.5\% between an observation altitude of \unit[1560]{m} 
above sea level and sea level itself. (This scatter arises from
slight clipping effects of the air-shower evolution at an observation altitude of \unit[1560]{m} above sea level and from
statistical uncertainties in the determination of the radiation energy
from the simulated radio-emission footprint.) The radiation energy directly reflects 
the calorimetric energy in the electromagnetic cascade of an extensive 
air shower, akin to an integral of the Gaisser-Hillas profile measured 
with fluorescence detectors. It constitutes a universal,
well-defined quantity that can be measured with radio detectors 
worldwide and can thus be compared directly between different 
experiments, as well as with theoretical predictions.

In this work, we measure the absolute value of the
radiation energy with the Auger Engineering Radio Array (AERA) 
\cite{ICRC2015JSchulz}, an array of radio detectors in the Pierre Auger Observatory \cite{Auger2014}. We then 
cross-calibrate our measurement with data taken with the baseline 
detectors of the Auger Observatory. The Observatory includes an array of water-Cherenkov
particle detectors covering an area of \unit[3,000]{km$^{2}$}. The 
atmosphere above the surface detector is monitored by fluorescence telescopes which provide an 
absolute calibration of the cosmic-ray energy scale \cite{Abraham2010239} with a systematic 
uncertainty of 16\% at \unit[$10^{17.5}$]{eV} and 14\% at energies 
$\geq 10^{18}${eV} \cite{EnergyScaleICRC2013}, reflecting the 
state-of-the-art in the determination of the absolute energy scale 
achieved to date. We thus use the accurate calibration of the 
energy scale of the Pierre Auger Observatory to relate the radiation 
energy to the cosmic-ray energy. The radiation energy can in turn be used to calibrate 
cosmic-ray radio detectors worldwide against the Auger energy scale. 
Finally, we provide a first comparison with predictions from 
first-principle calculations.

Details of the analysis presented here can be found in an 
accompanying publication \cite{AERAEnergyPRD}.

\emph{The energy content in the radio signal.}---\label{sec:energycontent}With the radio antennas of AERA, we 
continuously sample voltage traces arising from the measurement of the 
local electric field with antennas oriented along the geomagnetic
north-south and east-west directions. Upon a trigger from coincident radio pulses or 
external trigger information from other Auger detectors, the voltage traces are 
read out for off-line analysis \cite{OfflineRadio}. From these voltage 
traces, we reconstruct the electric field vector at the location of 
each radio detector as a function of 
time. Detector effects are carefully unfolded \cite{AERAEnergyPRD}. The 
uncertainty of the electric field amplitude between different 
measurements is dominated by temperature 
variations (4\%) and uncertainties of the antenna response pattern 
(5\%), and amounts to a total of 6.4\%. The uncertainty of
the absolute amplitude scale is dominated by the antenna
response (12.5\% \cite{PHDWeidenhaupt,AntennaPaper}) and the analog signal chain (6\%) and
amounts to a total of 14\%.

After digital processing (involving noise cleaning, up-sampling and 
enveloping), we identify radio pulses exceeding a suitable 
signal-to-noise threshold. We calculate the instantaneous Poynting flux
at each radio detector and integrate it over a time window of 
\unit[200]{ns} which is centered on the pulse maximum. The contribution of noise to the integral is estimated 
from data recorded before the arrival of the extensive air shower, and is subtracted from the integrated signal. The result 
of the time-integration corresponds to the energy deposited per area by air-shower radio signals at the locations of 
the individual radio detectors. We measure this \emph{energy fluence} 
in units of eV/m$^{2}$. Typical values are in the range of dozens of 
eV/m$^{2}$. The energy of a photon at our center-of-band frequency of 
\unit[55]{MHz} corresponds to \unit[$2.27 \times 10^{-7}$]{eV}. The number of received
photons is thus very high, illustrating that uncertainties from photon
statistics are negligible in radio detection of extensive air showers.

\begin{figure}[h!tb]
\includegraphics[width=0.48\textwidth]{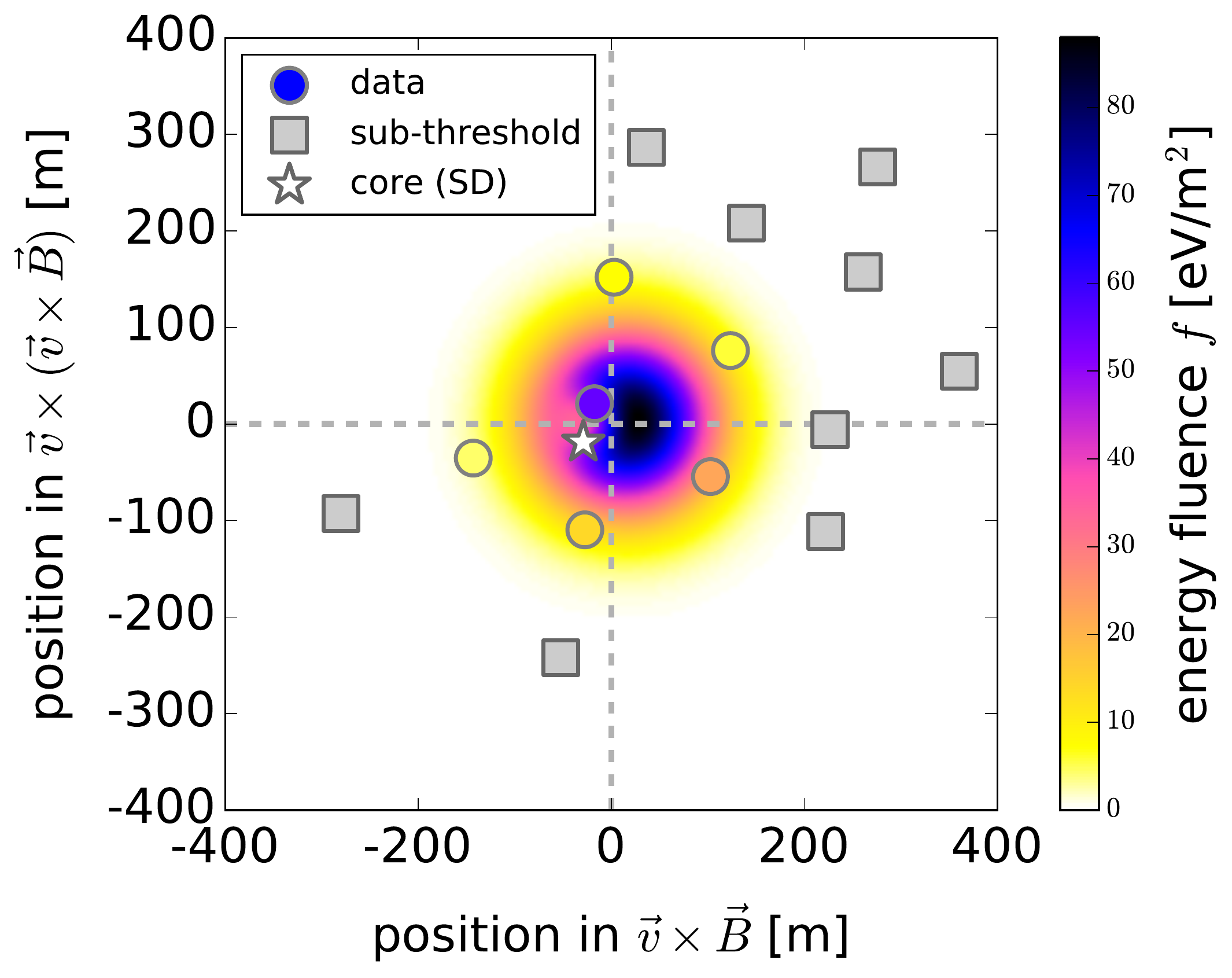}
\includegraphics[width=0.4\textwidth]{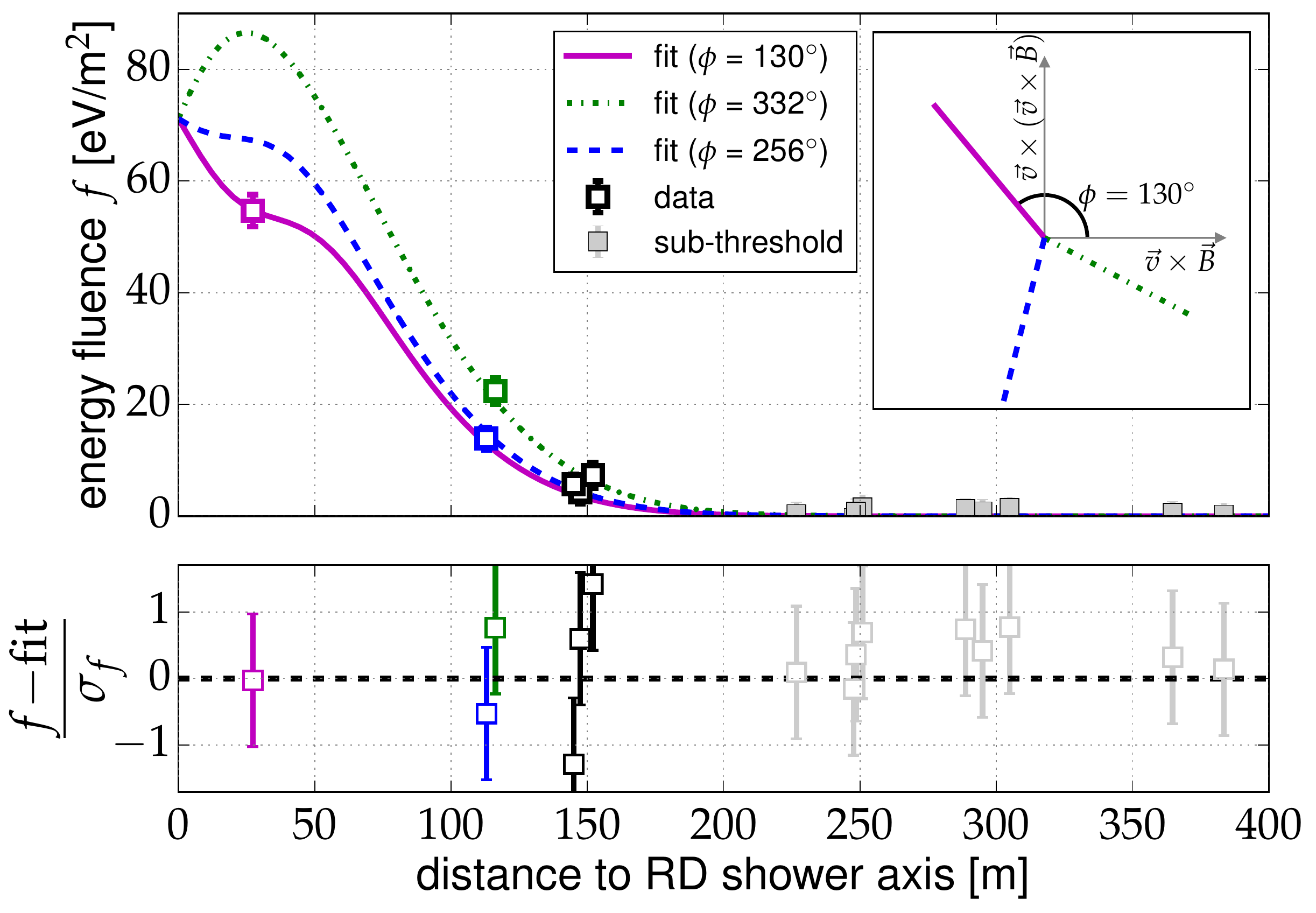}
\caption{Top: Energy fluence for an extensive air shower 
with an energy of \unit[$4.4 \times 10^{17}$]{eV}, and a zenith angle of 
25$^{\circ}$ as measured in individual AERA radio detectors (circles filled with color corresponding to the measured value)
and fitted with the azimuthally asymmetric, two-dimensional signal distribution function (background color). Both, radio detectors with a detected signal 
(\emph{data}) and below detection threshold (\emph{sub-threshold}) participate in the fit. 
The fit is performed in the plane perpendicular to the shower axis, with the $x$-axis oriented along 
the direction of the Lorentz force for charged particles 
propagating along the shower axis $\vec{v}$ in the geomagnetic field 
$\vec{B}$. The best-fitting impact point of the air shower is at the origin of the plot, 
slightly offset from the one reconstructed with the Auger surface 
detector (\emph{core (SD)}). Bottom: Representation of the same data 
and fitted two-dimensional signal distribution as a function of distance from the
shower axis. The colored and black squares denote the energy
fluence measurements, gray squares represent radio detectors with signal below threshold.
For the three data points with the highest energy fluence, the
one-dimensional projection of the two-dimensional signal distribution 
fit onto lines connecting the best-fitting impact point of the air shower with the corresponding radio detector 
positions is illustrated with colored lines. This demonstrates the 
azimuthal asymmetry and complexity of the two-dimensional signal distribution function. The inset figure illustrates 
the polar angles of the three projections. The distribution of the residuals (data versus fit) is shown
as well.\label{fig:2dldf}}
\end{figure}

The area illuminated by radio signals has a limited extent due to 
the forward-beamed nature of the radio emission. The local energy 
fluence at the radio detectors with an identified signal is fitted with a two-dimensional
distribution function of the signal \cite{LOFARLDF}, adapted to the 
observation altitude of AERA, which takes into
account azimuthal asymmetries arising from the superposition of
geomagnetic and charge-excess 
\cite{Askaryan1962,Marin2011,AERAPolarization} effects as well as
ring-shaped areas of enhanced emission caused by Cherenkov-like time compression due to the refractive index in 
the atmosphere \cite{DeVriesBergScholten2011,ZHAires2012}. During the fit procedure, spurious radio pulses not 
related to the extensive air shower are flagged and rejected by means of the 
signal polarization. In rare cases, flagging of spurious radio pulses can lead to rejection of a complete 
event. An example for the 
resulting fit is illustrated in Fig.\ \ref{fig:2dldf}. For radio events 
detected in three or four radio detectors,
the impact point of the shower axis used for the fit is fixed to the one reconstructed 
with the Auger surface detector. For radio events with signals in 
five or more radio detectors, the impact point is determined during the fit of 
the two-dimensional signal distribution function. 

After a successful fit of the signal distribution 
function we analytically integrate it over the plane perpendicular to 
the shower axis. The result is the 
total energy measured in the radio signal 
$E^\text{Auger}_{30-80\,\mathrm{MHz}}$ (in units of eV), the \emph{radiation energy}. This 
quantity does not depend on any characteristics of the detector 
except the finite measurement bandwidth from 30 to \unit[80]{MHz}. 
The superscript ``Auger'' emphasizes that this quantity applies to 
the geomagnetic field strength as present at the site of the Pierre 
Auger Observatory in southern Argentina.

\emph{Cross-calibration with the Auger energy 
scale.}---\label{sec:crosscalibration}To establish the relation 
between the radiation energy and the absolute energy scale of cosmic rays, we analyzed data from the first stage of AERA taken between April 2011 
and March 2013, when the array consisted of 24 radio detectors equipped 
with logarithmic-periodic dipole antennas \cite{AntennaPaper}. The 
signal distribution fit was applied to data pre-selected with standard Auger 
quality cuts for surface detector events measured with the 
\unit[750]{m} grid of the array. We allowed a maximum zenith angle of 
$55^{\circ}$ and required an energy of at least \unit[10$^{17}$]{eV}. 
This resulted in a data set with 126 events.

For each of these events, the cosmic-ray energy $E_{\mathrm{CR}}$ as reconstructed with the Auger 
surface detector \cite{AUGERICRC2011_Maris} is available. We stress 
that the energy reconstruction of the surface detector has been calibrated with the 
calorimetric energy measurement of the fluorescence detector using a 
subset of events measured with both detectors simultaneously. Due to the dominance of geomagnetic 
radio emission \cite{LOPES2005,CodalemaGeoMag,AERAPolarization} and 
the scaling of its amplitude with the magnitude of the Lorentz force, the radiation 
energy scales with $\sin^2(\alpha)$, where $\alpha$ denotes the angle between the air-shower
axis and the geomagnetic-field axis. We thus normalize the 
radiation energy for perpendicular incidence with respect to the geomagnetic field by dividing it by $\sin^2(\alpha)$. This normalization is valid for all incoming 
directions of cosmic rays except for a small region around the 
geomagnetic-field axis. In particular, it is valid for all events in the data set 
presented here.

\begin{figure}
\includegraphics[width=0.5\textwidth]{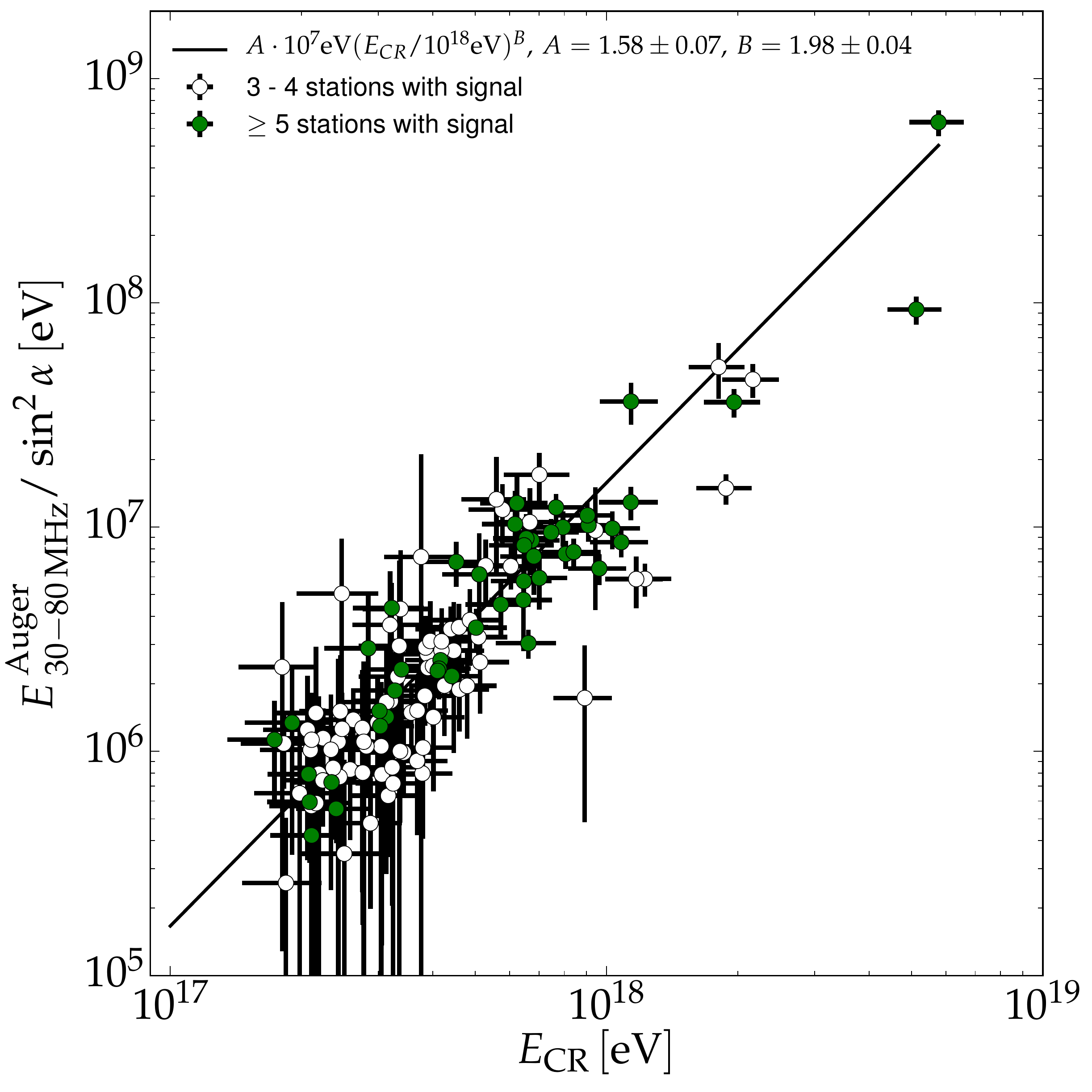}
\caption{Correlation between the normalized radiation energy and the 
cosmic-ray energy $E_{\text{CR}}$
as determined by the Auger surface detector. Open circles represent air showers with radio signals 
detected in three or four radio detectors. Filled circles 
denote showers with five or more detected radio signals.\label{fig:crosscalibration}}
\end{figure}

In Fig.\ \ref{fig:crosscalibration}, the value of $E^\text{Auger}_{30-80\,\mathrm{MHz}}/\sin^{2}(\alpha)$ for 
each measured air shower is plotted as a function of the cosmic-ray energy measured 
with the Auger surface detector. A log-likelihood fit taking into account threshold effects, 
measurement uncertainties and the steeply falling cosmic-ray energy 
spectrum \cite{Dembinski2015} shows that the data can be described well 
with the power law
\begin{equation}
\label{eq:calibration_sradio}
E^\text{Auger}_{30-80\,\mathrm{MHz}}/\sin^{2}(\alpha) = A \times \unit[10^7]{eV} 
\,(E_\text{CR}/\unit[10^{18}]{eV})^B.
\end{equation}
The result of the fit yields $A = 1.58 \pm 0.07$ and $B = 
1.98 \pm 0.04$. For a cosmic ray with an energy of \unit[1]{EeV} arriving 
perpendicularly to the Earth's magnetic field at the Pierre Auger 
Observatory, the radiation energy thus amounts to 
\unit[15.8]{MeV}, a minute fraction of the energy of the primary 
particle. The observed quadratic scaling is expected for coherent radio emission, for which amplitudes scale linearly and thus the radiated energy scales quadratically.

Taking into account the energy- and zenith-dependent uncertainty of 
$E_\text{CR}$,
the resolution of $E^\text{Auger}_{30-80\,\mathrm{MHz}}/\sin^{2}(\alpha)$ is determined 
from the scatter of points in Fig.\ \ref{fig:crosscalibration}. It amounts to 22\% for the full data set. Performing this analysis for the high-quality subset of events 
with a successful radio detection in at least five radio detectors 
yields a resolution of 17\%.

The value of $A$ reported here applies for a cosmic-ray shower with 
an energy of \unit[1]{EeV} evolving in a geomagnetic field 
with a strength of \unit[0.24]{G}, as present at the site of the Pierre Auger 
Observatory. With dedicated simulations we confirmed that the 
radiation energy is only marginally influenced by the charge-excess 
contribution (at the level of 2\% for showers arriving perpendicular 
to the magnetic field at the Pierre Auger site, less for stronger geomagnetic fields).
Hence, a normalization with the field strength of the geomagnetic field is possible and yields:
\begin{eqnarray}
 E_{30-80\,{\mathrm{MHz}}} &=& \left(15.8~ \pm 0.7\,\mathrm{(stat)} \pm 6.7 
\,\mathrm{(sys)}\right)\,\unit{MeV} \nonumber \\ &\times& \left(\sin\alpha 
\,\frac{E_\text{CR}}{\unit[10^{18}]{eV}} \, \frac{B_\mathrm{Earth}}{\unit[0.24]{G}} \right)^2.
\end{eqnarray}
$E_\mathrm{30-80 MHz}$ can be used by radio detectors worldwide for 
cross-calibration of the energy scale, except for experiments
deployed at high altitude where part of the radio emission 
is clipped when the shower reaches the ground before radiating the 
bulk of its radio emission. The frequency window from $\sim\,$30 to 
\unit[$\sim\,$80]{MHz} is shared by many radio detectors 
\cite{LOPES2005,Codalema2006,LOFAREnergy,Kostunin201489}: below 
\unit[30]{MHz} atmospheric noise and transmitters in the short-wave 
band dominate, above \unit[80]{MHz} coherence diminishes and the FM-band interferes with 
the measurement. Possible second-order effects arising in the 
determination of $E_\mathrm{30-80 MHz}$, e.g., due to shower geometry, should be addressed in a follow-up 
analysis because they could lead to further improvements. The systematic uncertainty of $E_\mathrm{30-80 MHz}$ quoted 
here arises from the quadratic sum of the systematic uncertainty on the energy scale of the 
Pierre Auger Observatory (16\% at \unit[$10^{17.5}$]{eV}, propagated from the fluorescence 
detector to the surface detector) and the uncertainty on the radio-electric field
amplitude measurement (14\%). These two contributions amount to uncertainties of 5.1 and \unit[4.4]{MeV}
in the measurement of the radiation energy at \unit[1]{EeV}, respectively. We note that the systematic uncertainty in the determination of 
the cosmic-ray energy from radio measurements is half of that of $E_\mathrm{30-80 MHz}$, as 
the cosmic-ray energy scales with the square root of the radiation 
energy.

\emph{Comparison with first-principle 
calculations.}---\label{sec:theorycomparison}In addition to a \emph{cross-calibration} of techniques and experiments against each other, the 
radiation energy can also be used for an \emph{independent determination}
of the absolute energy scale of cosmic-ray observatories. Sophisticated Monte Carlo simulations \cite{ZHAires2012,CoREAS2013,Marin2012733} 
provide a quantitative prediction of the radiation energy based on
first-principle calculations combining classical electrodynamics with the well-established properties of the 
electromagnetic cascade in extensive air showers. A direct comparison of the 
predicted and measured radiation energies can thus be used for an 
absolute determination of the energy scale of cosmic-ray detectors.

We have evaluated the radiation energy at a cosmic-ray energy of 
\unit[1]{EeV} using the typical zenith angle of our event sample of 37$^{\circ}$
and a geomagnetic field strength of \unit[0.24]{G} with the two available 
full Monte Carlo simulation codes CoREAS \cite{CoREAS2013} and ZHAireS 
\cite{ZHAires2012}. The predicted values for the radiation energy 
amount to \unit[11.9]{MeV} and \unit[11.3]{MeV}, respectively.
Both predictions are thus in agreement with our measurement 
within the quoted uncertainties. Further work will be undertaken to 
better understand and minimize experimental and theoretical systematic uncertainties.

\emph{Conclusions.}---\label{sec:conclusions}We have 
measured the \emph{radiation energy} of
extensive air showers and have used it as an energy estimator directly reflecting the 
calorimetric energy in the electromagnetic cascade. Its value is 
\unit[$15.8 \pm 0.7\,\mathrm{(stat)} \pm 6.7\,\mathrm{(sys)}$]{MeV} in the frequency band from 30 to \unit[80]{MHz} for a cosmic ray with 
an energy of \unit[$10^{18}$]{eV} arriving perpendicularly to a 
magnetic field with a strength of \unit[0.24]{G}. The radiation energy
can be measured at any location that does not suffer from strong anthropogenic noise using 
moderately sized radio detector arrays. It can thus be used for an 
efficient cross-calibration of the energy scales of different experiments and 
detection techniques against each other, in particular against the 
well-established energy scale of the Pierre Auger Observatory. Our measurement
is in agreement with predictions from first-principle calculations.

\begin{acknowledgments}

\emph{Acknowledgments.}---The successful installation, commissioning, and operation of the Pierre Auger
Observatory would not have been possible without the strong commitment and
effort from the technical and administrative staff in Malarg\"ue. We are
very grateful to the following agencies and organizations for financial
support:

\begin{sloppypar}
Comisi\'on Nacional de Energ\'{\i}a At\'omica,
Agencia Nacional de Promoci\'on Cient\'{\i}fica y Tecnol\'ogica (ANPCyT),
Consejo Nacional de Investigaciones Cient\'{\i}ficas y T\'ecnicas (CONICET),
Gobierno de la Provincia de Mendoza,
Municipalidad de Malarg\"ue,
NDM Holdings and Valle Las Le\~nas, in gratitude for their continuing cooperation over land access,
Argentina;
the Australian Research Council (DP150101622);
Conselho Nacional de Desenvolvimento Cient\'{\i}fico e Tecnol\'ogico (CNPq), Financiadora de Estudos e Projetos (FINEP),
Funda\c{c}\~ao de Amparo \`a Pesquisa do Estado de Rio de Janeiro (FAPERJ),
S\~ao Paulo Research Foundation (FAPESP) Grants No.\ 2010/07359-6 and No.\ 1999/05404-3,
Minist\'erio de Ci\^encia e Tecnologia (MCT),
Brazil;
Grant No.\ MSMT-CR LG13007, No.\ 7AMB14AR005, and the Czech Science Foundation Grant No.\ 14-17501S,
Czech Republic;
Centre de Calcul IN2P3/CNRS, Centre National de la Recherche Scientifique (CNRS),
Conseil R\'egional Ile-de-France,
D\'epartement Physique Nucl\'eaire et Corpusculaire (PNC-IN2P3/CNRS),
D\'epartement Sciences de l'Univers (SDU-INSU/CNRS),
Institut Lagrange de Paris (ILP) Grant No.\ LABEX ANR-10-LABX-63,
within the Investissements d'Avenir Programme Grant No.\ ANR-11-IDEX-0004-02,
France;
Bundesministerium f\"ur Bildung und Forschung (BMBF),
Deutsche Forschungsgemeinschaft (DFG),
Finanzministerium Baden-W\"urttemberg,
Helmholtz Alliance for Astroparticle Physics (HAP),
Helmholtz-Gemeinschaft Deutscher Forschungszentren (HGF),
Ministerium f\"ur Wissenschaft und Forschung, Nordrhein Westfalen,
Ministerium f\"ur Wissenschaft, Forschung und Kunst, Baden-W\"urttemberg,
Germany;
Istituto Nazionale di Fisica Nucleare (INFN),
Istituto Nazionale di Astrofisica (INAF),
Ministero dell'Istruzione, dell'Universit\'a e della Ricerca (MIUR),
Gran Sasso Center for Astroparticle Physics (CFA),
CETEMPS Center of Excellence, Ministero degli Affari Esteri (MAE),
Italy;
Consejo Nacional de Ciencia y Tecnolog\'{\i}a (CONACYT),
Mexico;
Ministerie van Onderwijs, Cultuur en Wetenschap,
Nederlandse Organisatie voor Wetenschappelijk Onderzoek (NWO),
Stichting voor Fundamenteel Onderzoek der Materie (FOM),
Netherlands;
National Centre for Research and Development, Grants No.\ ERA-NET-ASPERA/01/11 and No.\ ERA-NET-ASPERA/02/11,
National Science Centre, Grants No.\ 2013/08/M/ST9/00322, No.\ 2013/08/M/ST9/00728 and No.\ HARMONIA 5 - 2013/10/M/ST9/00062,
Poland;
Portuguese national funds and FEDER funds within Programa Operacional Factores de Competitividade through Funda\c{c}\~ao para a Ci\^encia e a Tecnologia (COMPETE),
Portugal;
Romanian Authority for Scientific Research ANCS,
CNDI-UEFISCDI partnership projects Grants No.\ 20/2012 and No.\ 194/2012,
Grants No.\ 1/ASPERA2/2012 ERA-NET, No.\ PN-II-RU-PD-2011-3-0145-17 and No.\ PN-II-RU-PD-2011-3-0062,
the Minister of National Education,
Programme Space Technology and Advanced Research (STAR), Grant No.\ 83/2013,
Romania;
Slovenian Research Agency,
Slovenia;
Comunidad de Madrid,
FEDER funds,
Ministerio de Educaci\'on y Ciencia,
Xunta de Galicia,
European Community 7th Framework Program, Grant No.\ FP7-PEOPLE-2012-IEF-328826,
Spain;
Science and Technology Facilities Council,
United Kingdom;
Department of Energy, Contracts No.\ DE-AC02-07CH11359, No.\ DE-FR02-04ER41300, No.\ DE-FG02-99ER41107 and No.\ DE-SC0011689,
National Science Foundation, Grant No.\ 0450696,
The Grainger Foundation,
USA;
NAFOSTED,
Vietnam;
Marie Curie-IRSES/EPLANET,
European Particle Physics Latin American Network,
European Union 7th Framework Program, Grant No.\ PIRSES-2009-GA-246806;
and
UNESCO.
\end{sloppypar}
\end{acknowledgments}

%

\end{document}